\documentclass[12pt]{article}
\usepackage{float}
\usepackage{amsmath}
\usepackage{graphicx}

\makeatletter
\usepackage{axodraw,bbold}

\catcode`@=12
\makeatother
\begin{document}
\thispagestyle{empty}

\begin{flushright}UCRHEP-T401\\
 November 2005\end{flushright}

\vspace{0.5in}

\begin{center}\textbf{\LARGE Hexagonal SU(3) Unification and}\\
 \vspace{0.1in} 
 \textbf{\LARGE its Manifestation at the TeV Scale}\\
 \vspace{1.0in} \textbf{Qing-Hong Cao, Shao-Long Chen, Ernest Ma}\\
 \vspace{0.1in} \textsl{Physics Department, University of California,
Riverside, California 92521, USA}\\
 \vspace{0.2in} \textbf{G. Rajasekaran}\\
 \vspace{0.1in} \textsl{Institute of Mathematical Sciences, Chennai
(Madras) 600113, India}\\
 \vspace{1.0in}\end{center}

\begin{abstract}
\ We consider $SU(3)_{C}\times SU(2)_{AL}\times SU(2)_{BL}\times U(1)_{Y}$
as the low-energy subgroup of supersymmetric $SU(3)^{6}$ unification.
This may imply small deviations from quark-lepton universality at the
TeV scale, as allowed by neutron-decay data. New particles are predicted
with specific properties. We discuss in particular the new heavy gauge 
bosons corresponding to $SU(2)_{AL} \times SU(2)_{BL} \to SU(2)_L$.
\end{abstract}
\newpage
\baselineskip 24pt

\section{Hexagonal SU(3) Model}

The extension from $SU(3)_{C}\times SU(3)_{L}\times SU(3)_{R}$ trinification
\cite{trini} to $SU(3)^{6}$ unification \cite{hex6} allows for
the natural anomaly-free implementation of chiral color \cite{c_clr}
and quark-lepton nonuniversality \cite{q_1,lm03} at the TeV scale.
In view of the fact that there is an experimental hint \cite{ndecay}
of the latter, but not the former, we explore the possibility that
the low-energy reduction of hexagonal $SU(3)$ unification is actually
$SU(3)_{C}\times SU(2)_{AL}\times SU(2)_{BL}\times U(1)_{Y}$ at the
TeV scale, where quarks couple to $SU(2)_{AL}$, but leptons may choose
either $SU(2)_{AL}$ or $SU(2)_{BL}$ or both, and the $SU(2)_{L}$
of the Standard Model (SM) is the diagonal subgroup of $SU(2)_{AL}\times 
SU(2)_{BL}$ \cite{ln05}. We show how supersymmetric unification at around 
$10^{16}$ GeV may be maintained with a suitable choice of new particle content
at the TeV scale and discuss their phenomenological consequences.\\

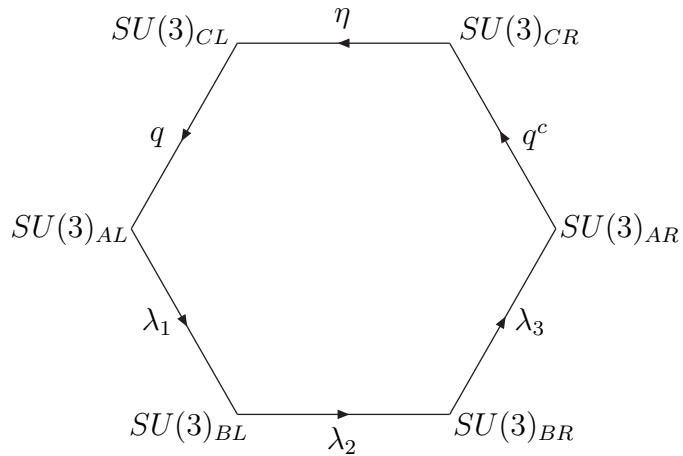
\begin{figure}[H]
\begin{center}\begin{picture}(270,150)(0,0)

\ArrowLine(100,0)(180,0)
\ArrowLine(60,70)(100,0)
\ArrowLine(100,140)(60,70)
\ArrowLine(180,140)(100,140)
\ArrowLine(220,70)(180,140)
\ArrowLine(180,0)(220,70)

\Text(70,105)[]{\(q\)}            \Text(75,145)[]{\(SU(3)_{CL}\)}
\Text(213,105)[]{\(q^c\)}         \Text(207,145)[]{\(SU(3)_{CR}\)}
\Text(70,35)[]{\(\lambda_1\)}             \Text(37,70)[]{\(SU(3)_{AL}\)}
\Text(211,35)[]{\(\lambda_3\)}          \Text(245,70)[]{\(SU(3)_{AR}\)}
\Text(140,-10)[]{\(\lambda_2\)}     \Text(82,-5)[]{\(SU(3)_{BL}\)}
\Text(140,150)[]{\(\eta\)}        \Text(205,-5)[]{\(SU(3)_{BR}\)}

\end{picture}\end{center}

\caption{Moose diagram of quarks and leptons in $[SU(3)]^{6}$.}
\end{figure}

We start with the supersymmetric $SU(3)^{6}$ model of Ref.~\cite{hex6}.
Under the gauge group $SU(3)_{CL}\times SU(3)_{AL}\times SU(3)_{BL}\times 
SU(3)_{BR}\times SU(3)_{AR}\times SU(3)_{CR}$, the six links of the 
{}``moose'' chain \cite{moose} are given by
\begin{eqnarray}
q & \sim & (3,3^{*},1,1,1,1),\\
\lambda_{1} & \sim & (1,3,3^{*},1,1,1),\\
\lambda_{2} & \sim & (1,1,3,3^{*},1,1),\\
\lambda_{3} & \sim & (1,1,1,3,3^{*},1),\\
q^{c} & \sim & (1,1,1,1,3,3^{*}),\\
\eta & \sim & (3^{*},1,1,1,1,3),\end{eqnarray}
 as shown in Fig.~1. The electric charge is embedded into $SU(3)^{6}$
according to \begin{equation}
Q=(I_{3})_{AL}+(I_{3})_{AR}-{\frac{1}{2}}Y_{AL}-{\frac{1}{2}}Y_{AR}+
(I_{3})_{BL}+(I_{3})_{BR}-{\frac{1}{2}}Y_{BL}-{\frac{1}{2}}Y_{BR}.
\end{equation}
Using the notation where the rows denote $(I_{3},Y)=(1/2,1/3),(-1/2,1/3),
(0,-2/3)$ and the columns denote $(I_{3},Y)=(-1/2,-1/3),(1/2,-1/3),(0,2/3)$,
the particle content of this model is given in matrix form as 
\begin{equation}
q=\begin{pmatrix}d & u & h \cr d & u & h \cr d & u & h\end{pmatrix},~~~
q^{c}=\begin{pmatrix}d^{c} & d^c & d^c \cr u^c & u^c & u^c \cr h^c & h^c & h^c
\end{pmatrix},~~~\lambda_{i}=\begin{pmatrix}N_{i} & E^c_i & \nu_i \cr E_i & 
N^c_i & e_i \cr \nu^c_i & e^c_i & S_i\end{pmatrix},
\end{equation}
and all the components of $\eta$ are neutral. As shown in Ref.~{[}2{]},
this embedding of electric charge yields the canonical value of 3/8
for $\sin^{2}\theta_{W}$ at the unification scale $M_{U}$. 

Whereas the quarks are unambiguously assigned in Eq.~(8), the leptons
are not. The left-handed doublets may be any linear combination of
$(\nu_{1},e_{1})$ and $(\nu_{2},e_{2})$, while the right-handed
doublets may be any linear combination of $(\nu_{2}^{c},e_{2}^{c})$
and $(\nu_{3}^{c},e_{3}^{c})$. We will see later exactly how this
works. Note that if $SU(3)^{6}$ collapses to $SU(3)^{3}$ already
at $M_{U}$, the leptons would then be unambigously assigned to $\lambda_{2}$. 

\section{Gauge Coupling Unification}

Above $M_{U}$, the six gauge couplings are assumed equal, maintained
for example with a discrete $Z_{6}$ symmetry. At $M_{U}$, $SU(3)^{6}$
is assumed broken down to \[
SU(3)_{C}\times SU(2)_{AL}\times SU(2)_{BL}\times U(1)_{Y}\]
 with the boundary conditions 
\begin{eqnarray}
{\frac{1}{\alpha_{C}(M_{U})}} & = & 
{\frac{1}{\alpha_{CL}(M_{U})}}+{\frac{1}{\alpha_{CR}(M_{U})}}={\frac{2}{\alpha_{U}}},\\
{\frac{1}{\alpha_{AL}(M_{U})}} & = & 
{\frac{1}{\alpha_{BL}(M_{U})}}={\frac{1}{\alpha_{U}}},\\
{\frac{3}{5\alpha_{Y}(M_{U})}} & = & 
{\frac{2}{\alpha_{U}}}.
\end{eqnarray}
 At $M_{S}$, supersymmetry is assumed broken, together with the breaking
of $SU(2)_{AL}\times SU(2)_{BL}$ to $SU(2)_{L}$ with the boundary
condition 
\begin{equation}
{\frac{1}{\alpha_{L}(M_{S})}}={\frac{1}{\alpha_{AL}(M_{S})}}+{\frac{1}{\alpha_{BL}(M_{S})}}.
\end{equation}

Consider now the one-loop renormalization-group equations governing
the evolution of the gauge couplings with mass scale: 
\begin{equation}
{\frac{1}{\alpha_{i}(M_{1})}}-{\frac{1}{\alpha_{i}(M_{2})}}={\frac{b_{i}}{2\pi}}\ln{\frac{M_{2}}{M_{1}}},
\end{equation}
where $\alpha_{i}=g_{i}^{2}/4\pi$ and the numbers $b_{i}$ are determined
by the particle content of the model between $M_{1}$ and $M_{2}$.
Below $M_{S}$, we assume the particle content of the SM, but with
two Higgs doublets, i.e. 
\begin{eqnarray}
SU(3)_{C}: &  & b_{C}=-11+(4/3)N_{f}=-7,\\
SU(2)_{L}: &  & b_{L}=-22/3+(4/3)N_{f}+1/3=-3,\\
U(1)_{Y}: &  & b_{Y}=(20/9)N_{f}+1/3=7,
\end{eqnarray}
where $N_{f}=3$ is the number of families. Above $M_{S}$, the gauge
group becomes $SU(3)_{C}\times SU(2)_{AL}\times SU(2)_{BL}\times U(1)_{Y}$
with the following minimum particle content for each family: 
\begin{eqnarray}
 &  & (u,d)\sim(3,2,1,1/6),~~u^{c}\sim(3^{*},1,1,-2/3),~~d^{c}\sim(3^{*},1,1,1/3),\\
 &  & (\nu_{1},e_{1})\sim(1,2,1,-1/2),~~(e_{1}^{c},\nu_{1}^{c})\sim(1,1,2,1/2),\\
 &  & (\nu_{2},e_{2})\sim(1,1,2,-1/2),~~e^{c}\sim(1,1,1,1).
\end{eqnarray}
The $SU(2)_{AL}$ anomalies are canceled between $(u,d)$ and $(\nu_{1},e_{1})$,
whereas the $SU(2)_{BL}$ anomalies are canceled between $(\nu_{2},e_{2})$
and $(e_{1}^{c},\nu_{1}^{c})$. In addition, we assume the appearance
of one copy of $\eta\sim(8,1,1,0)$, two copies of $(N_{1},E_{1};E_{1}^{c},N_{1}^{c})\sim(1,2,2,0)$,
one copy of $(N_{2},E_{2};E_{2}^{c},N_{2}^{c})\sim(1,1,2,\mp1/2)$,
one copy of $(N_{4},E_{4};E_{4}^{c},N_{4}^{c})\sim(1,2,1,\mp1/2)$,
and one copy of $(E_{5}^{c},N_{5};N_{5}^{c},E_{5})\sim(1,2,1,\pm1/2)$,
where $\lambda_{4}\sim(1,3,1,1,3^{*},1)$ and $\lambda_{5}\sim(1,3^{*},1,1,3,1)$
are extra supermultiplets to be discussed later. 

The corresponding $b_{i}$'s are then given by 
\begin{eqnarray}
SU(3)_{C}: &  & b_{C}=-9+2N_{f}+3=0,\\
SU(2)_{AL}: &  & b_{AL}=-6+2N_{f}+4=4,\\
SU(2)_{BL}: &  & b_{BL}=-6+N_{f}+3=0,\\
U(1)_{Y}: &  & b_{Y}=(13/3)N_{f}+3=16,
\end{eqnarray}
Using Eqs.~(9) to (16), these imply the following two constraints
\cite{hex6}: 
\begin{eqnarray}
{\frac{1}{\alpha_{C}(M_{Z})}} & = & {\frac{3}{7}}\left[{\frac{4}{\alpha_{L}(M_{Z})}}-{\frac{1}{\alpha_{Y}(M_{Z})}}\right]+{\frac{4}{7\pi}}\ln{\frac{M_{S}}{M_{Z}}},\\
\ln{\frac{M_{U}}{M_{Z}}} & = & {\frac{\pi}{14}}\left[{\frac{3}{\alpha_{Y}(M_{Z})}}-{\frac{5}{\alpha_{L}(M_{Z})}}\right]+{\frac{2}{7}}\ln{\frac{M_{S}}{M_{Z}}}.
\end{eqnarray}
Using the input \cite{pdg}
\begin{eqnarray}
\alpha_{L}(M_{Z}) & = & (\sqrt{2}/\pi)G_{F}M_{W}^{2}=0.0340,\\
\alpha_{Y}(M_{Z}) & = & \alpha_{L}(M_{Z})\tan^{2}\theta_{W}=0.0102,
\end{eqnarray}
and 
\begin{equation}
0.115<\alpha_{C}(M_{Z})<0.119,
\end{equation}
we find 
\begin{equation}
450~{\textrm{GeV}}>M_{S}>M_{Z},
\end{equation}
and 
\begin{equation}
1.2\times10^{16}~{\textrm{GeV}}<M_{U}<2.0\times10^{16}~{\textrm{GeV}}.
\end{equation}
These are certainly acceptable values for new particles below the
TeV scale and the proper suppression of proton decay. 

\section{Quarks, Leptons, and Other Particles}

Quark masses come from the Yukawa couplings $u^{c}(uN_{4}^{c}-dE_{4}^{c})$
and $d^{c}(uE_{4}-dN_{4})$ which originate from the invariant dimension-four
term $q^{c}\eta q\lambda_{4}$ term in the $SU(3)^{6}$ superpotential.
One of the $\eta$ supermultiplets is assumed to have superheavy vacuum
expectation values $\langle\eta_{11}\rangle=\langle\eta_{22}\rangle=\langle\eta_{33}\rangle$
which break $SU(3)_{CL}\times SU(3)_{CR}$ to $SU(3)_{C}$ at $M_{U}$.
Thus $q$ and $q^{c}$ become triplets and antitriplets respectively
under $SU(3)_{C}$, and an effective $q^{c}q\lambda_{4}$ term is
generated. 

To preserve the discrete $Z_{6}$ symmetry, $\lambda_{4}\sim(1,3,1,1,3^{*},1)$
should be accompanied by $Q^{c}\sim(1,1,3,1,1,3^{*})$, $\bar{Q}\sim(3^{*},1,,1,3,1,1)$,
$\lambda_{5}\sim(1,3^{*},1,1,3,1)$, $\bar{Q}^{c}\sim(1,1,3^{*},1,1,3)$,
and $Q\sim(3,1,1,3^{*},1,1)$. It is clear that $Q\bar{Q}$ and $Q^{c}\bar{Q}^{c}$,
as well as $\lambda_{4}\lambda_{5}$ are invariants so that all these
particles are naturally superheavy. However, $\lambda_{4}^{3}$ and
$\lambda_{5}^{3}$ are also invariants, so some of the components
of $\lambda_{4}$ and $\lambda_{5}$ may be fine-tuned to be light. 

At $M_{S}$, one of the $(1,2,2,0)$ bidoublets is assumed to have
vacuum expectation values $\langle N_{1}\rangle=\langle N_{1}^{c}\rangle$
which break $SU(2)_{AL}\times SU(2)_{BL}$ to $SU(2)_{L}$. From the
invariant $\lambda_{1}^{3}$ term, $(\nu_{1},e_{1})$ will then pair
with $(e_{1}^{c},\nu_{1}^{c})$ to form a vector doublet under $SU(2)_{L}$,
and from the invariant $\lambda_{2}^{3}$ term, $(\nu_{2},e_{2})$
will couple to $e_{2}^{c}$ through $(N_{2},E_{2})$ to become the
SM leptons, as in $SU(3)^{3}$ trinification. This is also the canonical
case of quark-lepton nonuniversality \cite{lm03} because quarks couple
to $SU(2)_{AL}$ and leptons couple to $SU(2)_{BL}$. 

However, there is also the $\lambda_{1}\lambda_{2}\lambda_{3}\lambda_{5}$
term in the $SU(3)^{6}$ superpotential. One of the $\lambda_{3}$
supermultiplets is assumed to have superheavy vacuum expectation values
$\langle N_{3}\rangle=\langle N_{3}^{c}\rangle=\langle S_{3}\rangle$
which break $SU(3)_{AR}\times SU(3)_{BR}$ to $SU(3)_{R}$ at $M_{U}$.
Thus $(\nu_{1},e_{1})$ may couple to $e_{2}^{c}$ through $(N_{5}^{c},E_{5})$.
At the same time, one of the $\lambda_{1}$ supermultiplets is assumed
to have a superheavy vacuum expectation value $\langle S_{1}\rangle$
which breaks $SU(3)_{AL}\times SU(3)_{BL}$ to $SU(2)_{AL}\times SU(2)_{BL}\times U(1)_{YL}$
at $M_{U}$. Thus $(\nu_{2},e_{2})$ may also couple to $e_{3}^{c}$
through $(N_{5}^{c},E_{5})$. In either case, the lepton doublet and
the antilepton singlet would be in different $(3,3^{*})$ reprsentations,
as in two previously proposed models \cite{bmw,m05}. To break $SU(3)_{R}\times U(1)_{YL}$
to $U(1)_{Y}$, we assume superheavy vacuum expectation values $\langle\nu_{3}\rangle$
and $\langle S_{2}\rangle$ as well. As shown in Ref.~\cite{m05},
having $(\nu,e)$ and $(e^{c},\nu^{c})$ in separate $(3,3^{*})$
representations allows $\nu^{c}$ to acquire a large Majorana mass,
thereby realizing the canonical seesaw mechanism for very small Majorana
neutrino masses. This argues for the scenario where the SM leptons
are not exclusively from $\lambda_{2}$ as in the original $SU(3)^{6}$
model \cite{hex6}. 

The new particles at $M_{S}$ all have $SU(2)_{L}\times U(1)_{Y}$
invariant masses and do not contribute significantly to the $S,T,U$
oblique parameters, thereby preserving the excellent agreement of
the SM with current precision electroweak measurements \cite{pdg}.
The $SU(3)_{C}$ octet $\eta$ decays in one loop to two gluons, and
should be detected at the Large Hadron Collider (LHC). The
$SU(3)_{C}$ singlets interact with one another through the terms
$\lambda_{1}\lambda_{2}\lambda_{5}$ and $\lambda_{4}\lambda_{5}$,
which allow them to decay into SM particles, such as leptons and quarks
as well as $W$ and $Z$ bosons. 

In the Minimal Supersymmetric Standard Model, the leptonic doublet
has to be distinguished from the Higgs doublet of the same hypercharge
by $R$-parity to guarantee the existence of a stable particle, the
Lightest Supersymmetric Particle (LSP), as a candidate for dark matter.
Here the Higgs superfields are all bidoublets and leptons doublets,
so they are already distinguished by the structure of the theory and
an effective $R$-parity exists automatically. 

\section{New Gauge Bosons at the TeV Scale}

The salient feature of this model is of course the appearance of a
second set of weak gauge bosons corresponding to the breaking of $SU(2)_{AL}\times SU(2)_{BL}$
to the $SU(2)_{L}$ of the SM. As a result, the left-handed quark
doublet $(u,d)$ couples to \[
g_{L}W+{\frac{g_{A}^{2}}{\sqrt{g_{A}^{2}+g_{B}^{2}}}}W',\]
 and the left-handed lepton doublet $(\nu,e)$ couples to \[
g_{L}W+{\frac{g_{A}^{2}\cos^{2}\theta-g_{B}^{2}\sin^{2}\theta}{\sqrt{g_{A}^{2}+g_{B}^{2}}}}W',\]
 where $g_{L}^{-2}=g_{A}^{-2}+g_{B}^{-2}$ and the SM set of $SU(2)_{L}$
gauge bosons $W$ and their orthogonal combinations $W'$ are given
by 
\begin{equation}
W={\frac{g_{B}W_{A}+g_{A}W_{B}}{\sqrt{g_{A}^{2}+g_{B}^{2}}}},
~~~W'={\frac{g_{A}W_{A}-g_{B}W_{B}}{\sqrt{g_{A}^{2}+g_{B}^{2}}}},
\end{equation}
with 
\begin{equation}
\begin{pmatrix}\nu\cr e\end{pmatrix}=
\cos\theta\begin{pmatrix}\nu_{1}\cr e_{1}\end{pmatrix}+\sin\theta\begin{pmatrix}\nu_{2}\cr e_{2}\end{pmatrix}.
\end{equation}
 If $\theta=0$, then quarks and leptons interact identically with
$W'$ as well as $W$. If $\theta=\pi/2$, then we have the canonical
case of quark-lepton nonuniversality \cite{lm03}.

\subsection{$W'$ coupling}

In general, $W$ can mix with $W'$. For illustration, let us consider
the simpler case of no mixing in which the coupling of $q$-$W'$-$q'$
is $$ig_{L}\gamma^{\mu}P_{L}g_{W'qq},$$
and the coupling of $\ell$-$W'$-$\ell'$ is \[
ig_{L}\gamma^{\mu}P_{L}g_{W'\ell\ell}.\]
Here, $g_{L}$ is the SM $SU(2)_L$ coupling, $P_L = (1-\gamma_5)/2$, and 
the coefficients $g_{W'qq}$ and $g_{W'\ell\ell}$ are defined as follows:
\begin{eqnarray}
g_{W'qq} & = & \frac{g_{A}}{g_{B}},\nonumber \\
g_{W'\ell\ell} & = & \frac{g_{A}}{g_{B}}\cos^{2}\theta-\frac{g_{B}}{g_{A}}\sin^{2}\theta.\label{eq:effective_coupling}
\end{eqnarray}
The effective Fermi constant $G_{F}/\sqrt{2}$ in nuclear beta decay is then given by 
\begin{eqnarray}
\left({\frac{G_{F}}{\sqrt{2}}}\right)_{q\ell} 
& = & 
\frac{g_{L}^{2}(M_{Z})}{8M_{W}^{2}}+\frac{g_{L}^{2}(M_{S})}{8M_{W'}^{2}}g_{W'qq}g_{W'\ell\ell}\nonumber \\
 & = & \frac{g_{L}^{2}(M_{Z})}{8M_{W}^{2}}\left[1+\frac{\alpha_{L}(M_{S})M_{W}^{2}}{\alpha_{L}(M_{Z})M_{W'}^{2}}\frac{g_{A}}{g_{B}}\left(\frac{g_{A}}{g_{B}}\cos^{2}\theta-\frac{g_{B}}{g_{A}}\sin^{2}\theta\right)\right],\label{eq:gfql}\end{eqnarray}
whereas that in pure leptonic decay is 
\begin{eqnarray}
\left({\frac{G_{F}}{\sqrt{2}}}\right)_{\ell\ell} & = & \frac{g_{L}^{2}(M_{Z})}{8M_{W}^{2}}+\frac{g_{L}^{2}(M_{S})}{8M_{W'}^{2}}g_{W'\ell\ell}g_{W'\ell\ell}\nonumber \\
 & = & \frac{g_{L}^{2}(M_{Z})}{8M_{W}^{2}}\left[1+\frac{\alpha_{L}(M_{S})M_{W}^{2}}{\alpha_{L}(M_{Z})M_{W'}^{2}}\left(\frac{g_{A}}{g_{B}}\cos^{2}\theta-\frac{g_{B}}{g_{A}}\sin^{2}\theta\right)^{2}\right].\label{eq:gfll}\end{eqnarray}
 Therefore, if $\tan^{2}\theta>g_{A}^{2}/g_{B}^{2}$, then $(G_{F})_{q\ell}
<(G_{F})_{\ell\ell}$
and the neutron-decay result can be understood \cite{lm03}. Furthermore,
if $|g_{W'\ell\ell}| << |g_{W'qq}|$,
then $(G_{F})_{\ell\ell}$ is very close to $G_{F}^{SM}$, and $(G_{F})_{q\ell}$
will be less than it by a small amount.

\begin{figure}[t]
\centering\includegraphics[%
  clip,
  scale=0.65]{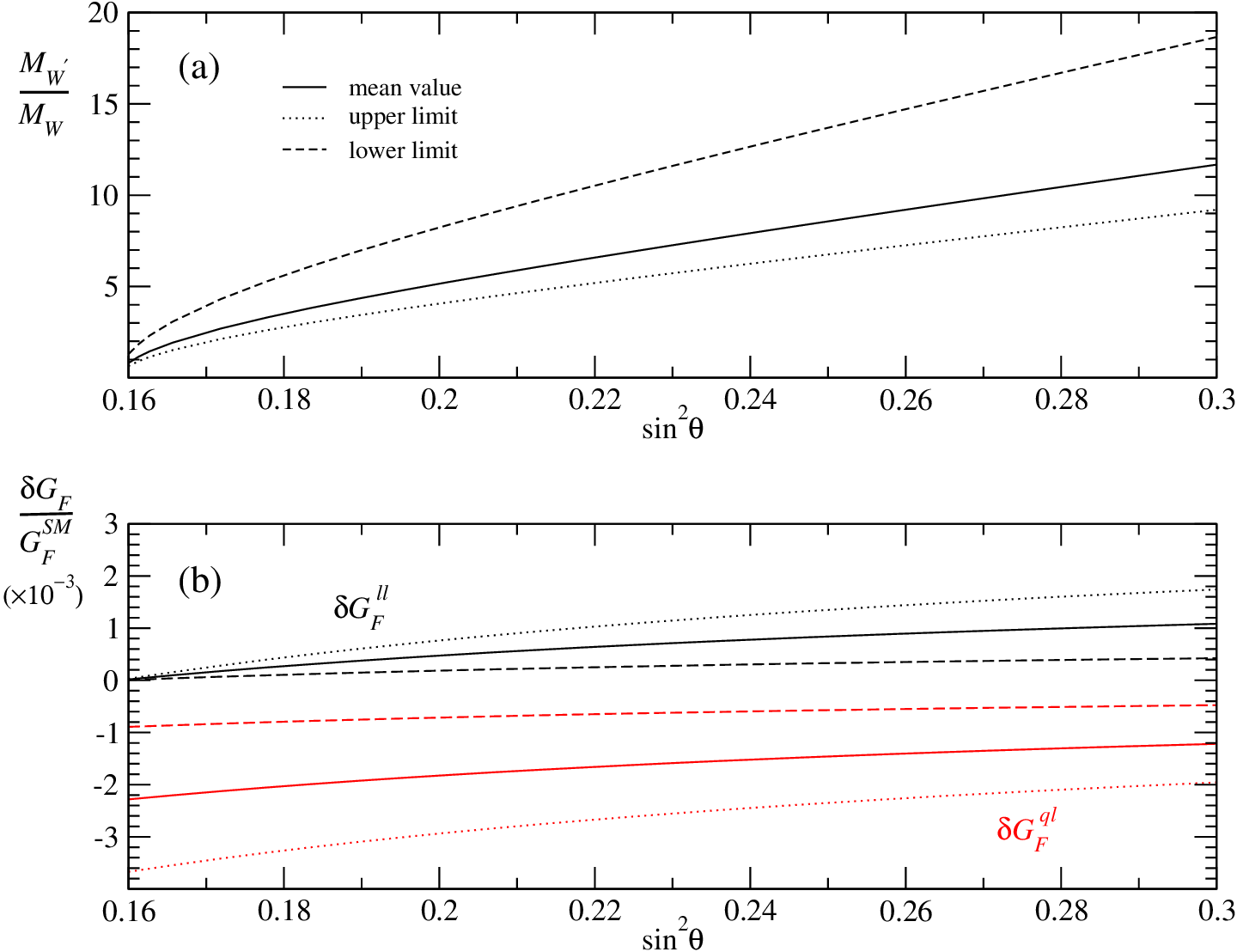}

\caption{(a) $M_{W'}/M_{W}$ as a function of $\sin^{2}\theta$, cf. Eq.~(\ref{eq:mwp-sth}).
(b) The corresponding deviations of $(G_{F})_{ql}$ and $(G_{F})_{ll}$
from $G_{F}^{SM}$ as functions of $\sin^{2}\theta$, cf. Eqs.~(\ref{eq:gfql})
and (\ref{eq:gfll}). The solid curve is obtained from the mean value
while the dotted and dashed curves are obtained from the upper
and lower values respectively.\label{fig:mwp-sth}}
\end{figure}

In general, $M_{W'}$ and $\sin\theta$ are independent parameters.
But in order to explain the neutron-decay result \cite{ndecay}, we should have 
\begin{equation}
1-\frac{(G_{F})_{q\ell}}{(G_{F})_{\ell\ell}}\simeq\frac{\alpha_{L}(M_{S})M_{W}^{2}}{\alpha_{L}(M_{Z})M_{W'}^{2}}g_{W'\ell\ell}\left(g_{W'\ell\ell}-g_{W'qq}\right)\simeq0.0023\pm0.0014.\label{eq:neutron-decay-result}
\end{equation}
Here we have used the latest value of $|V_{us}| = 0.2262(23)$ \cite{ckm2005} 
instead of the 2004 PDG value of 0.2200(26).  This reduces significantly the 
possible descrepancy of the neutron-decay result from universality.  
Using Eqs.~(21) and (22) as well as $M_{S}/M_{Z}=2.2$ and $M_{U}/M_{Z}=1.7\times10^{14}$
from $\alpha_{C}(M_{Z})=0.117$, we find $\alpha_{A}(M_{S})=0.040$
and $\alpha_{B}(M_{S})=0.212$. Hence we obtain the following relation
between $M_{W'}$ and $\sin\theta$: 
\begin{equation}
{\frac{M_{W}^{2}}{M_{W'}^{2}}}\sin^{2}\theta(\sin^{2}\theta-0.1587)\simeq3.11\pm1.89\times10^{-4}.\label{eq:mwp-sth}
\end{equation}
For illustration, we show $M_{W'}/M_{W}$ as a function of $\sin^{2}\theta$
in Fig.~\ref{fig:mwp-sth}(a). The solid curve is obtained from the
central value of the right-hand side of Eq.~(\ref{eq:neutron-decay-result}) while the dotted
and dashed curves are obtained from the upper and lower values
respectively.  For a given value of $\sin^2 \theta$, $M_{W'}$ lies within 
a range of values as shown.  Correspondingly, the deviations of 
$(G_F)_{q\ell}$ and $(G_F)_{\ell\ell}$ from $G_F^{SM}$ are also correlated 
with $\sin^2 \theta$. We present these deviations as functions of 
$\sin^{2}\theta$ in Fig.~\ref{fig:mwp-sth}(b).  Since $(G_{F})_{\ell\ell}$ 
has been measured very preicsely, smaller values of $\sin^2 \theta$ are 
preferred.

\begin{figure}[t]
\centering\includegraphics[%
  clip,
  scale=0.65]{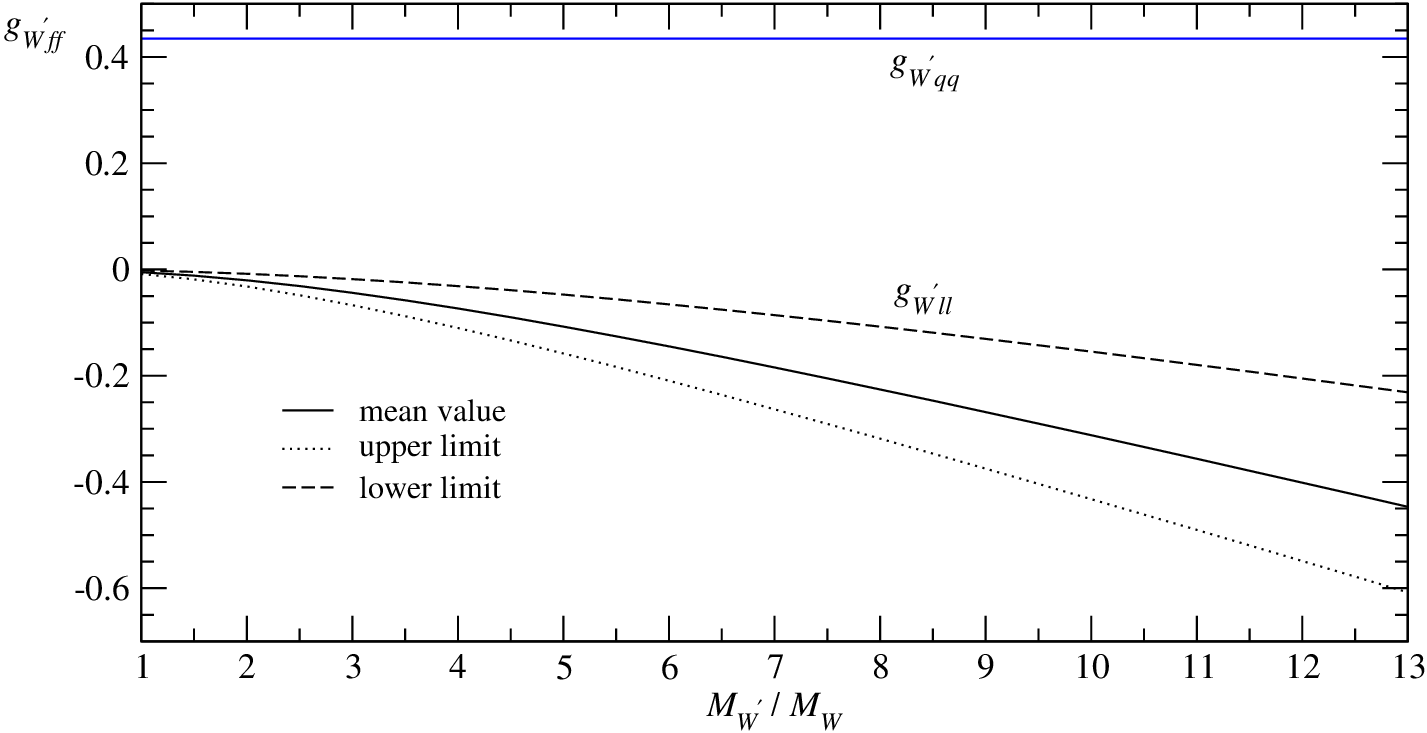}

\caption{The coupling strengths $g_{W'qq}$ and $g_{W'\ell\ell}$ as a function
of $M_{W'}/M_{W}$.\label{fig:coup-mwp}}
\end{figure}

In the following, we will choose the mass of the $W'$ boson as an input 
parameter rather than the mixing angle $\theta$.   Since the
effective coupling strength $g_{W'\ell\ell}$
is a function of $\sin^{2}\theta$, cf. Eq.~(\ref{eq:effective_coupling}),
it is also a function of $M_{W'}$.  Of course this dependence is not 
intrinsic to the model; it is simply due to the empirical constraint of 
Eq.~(\ref{eq:mwp-sth}).  For illustration, the effective
coupling strengths $g_{W'qq}$ and $g_{W'\ell\ell}$, as functions
of $M_{W'}$, are shown in Fig.~\ref{fig:coup-mwp}. Again,
the dotted curve is obtained from the upper limit and the dashed curve
from the lower limit. We note that both couplings are
suppressed compared to a SM-like coupling for which $g_{W'qq}=1$
and $g_{W'\ell\ell}=1.$ Furthermore, the magnitude of $g_{W'\ell\ell}$
is highly suppressed for a light $W'$ boson and grows graduately
with increasing $M_{W'}$. The difference between $g_{W'qq}$ and
$g_{W'\ell\ell}$ has a very important impact on the phenomenology
of $W'$ which will be addressed below.

\subsection{Decay of $W'$ boson}

Similar to the $W$ boson decay in the Standard Model, the $W'$ boson of this 
model can decay also into lepton pairs and quark pairs.  [Its decay into SM 
gauge bosons is negligible in the absence of mixing.] Taking into account 
the masses of the decay products, the $W'$ partial decay width is given by
\begin{eqnarray}
\Gamma\left(W'\to f\bar{f^{\prime}}\right) 
& = & 
N_{C}\frac{g^{2}M_{W'}}{48\pi}g_{W'ff}^{2}\lambda^{1/2}\left(1,\gamma_{f},\gamma_{f^{\prime}}\right)\nonumber \\
 & \times & \left[1-\frac{1}{2}\gamma_{f}-\frac{1}{2}\gamma_{f^{\prime}}-\frac{1}{2}\left(\gamma_{f}-\gamma_{f^{\prime}}\right)^2\right],
\end{eqnarray}
where $\gamma_{f}=m_{f}^{2}/M_{W'}^{2}$, $\gamma_{f^{\prime}}=m_{f^{\prime}}^{2}/M_{W'}^{2}$
and $\lambda(a,b,c)=a^{2}+b^{2}+c^{2}-2ab-2ac-2bc.$ Here, $N_{C}$
is the color factor of the fermion and $g_{W'ff}$ denotes either $g_{W'qq}$ 
or $g_{W'\ell\ell}$, as defined in Eq.~(\ref{eq:effective_coupling}).  
Of the hadronic modes, we need to consider only the decays $W'\to u\bar{d}$, 
$W'\to c\bar{s},$ and $W'\to t\bar{b}$ because  
\[ \left|V_{ud}\right|\approx\left|V_{cs}\right|\approx\left|V_{tb}\right|
\approx1\]
and all other CKM matrix elements are small. Since the $W'$ boson is very 
heavy, we can treat all
its decay products as massless particles except for the top quark.
The leptonic decay width of $W'$ boson can now be simplified as 
\begin{equation}
\Gamma\left(W'\to\ell\bar{\ell^{\prime}}\right)=\frac{g_{L}^{2}}{48\pi}M_{W'}g_{W'\ell\ell}^{2},
\end{equation}
where $\ell\ell^{\prime}=e\nu_{e},\mu\nu_{\mu},\tau\nu_{\nu}$.
If $M_{W'}<m_{t}$, the $W'$ boson can only decay into light quark
pairs,
\begin{equation}
\Gamma\left(W'\to q\bar{q^{\prime}}\right)=N_{C}\frac{g_{L}^{2}}{48\pi}M_{W'}g_{W'qq}^{2},
\end{equation}
where $qq^{\prime}=ud,\,\, cs.$  If $M_{W'}>m_{t}$, the $tb$
decay channel opens up and the partial decay width becomes
\begin{equation}
\Gamma\left(W'\to t\bar{b}\right)=N_{C}\frac{g_{L}^{2}}{48\pi}M_{W'}g_{W'qq}^{2}\left(1-\frac{3}{2}\gamma_{t}+\frac{1}{2}\gamma_{t}^{3}\right).
\end{equation}
Therefore, the total decay width of $W'$ is 
\begin{eqnarray}
\Gamma_{W'}^{tot}\left(M_{W'}<m_{t}\right) & = & \frac{g_{L}^{2}}{48\pi}M_{W'}\left(3g_{W'\ell\ell}^{2}+6g_{W'qq}^{2}\right),\\
\Gamma_{W'}^{tot}\left(M_{W'}>m_{t}\right) & = & \frac{g_{L}^{2}}{48\pi}M_{W'}\left[3g_{W'\ell\ell}^{2}+9g_{W'qq}^{2}\left(1-\frac{1}{2}\gamma_{t}+\frac{1}{6}\gamma_{t}^{3}\right)\right].
\end{eqnarray}
\begin{figure}[t]
\centering
\includegraphics[clip,scale=0.65]{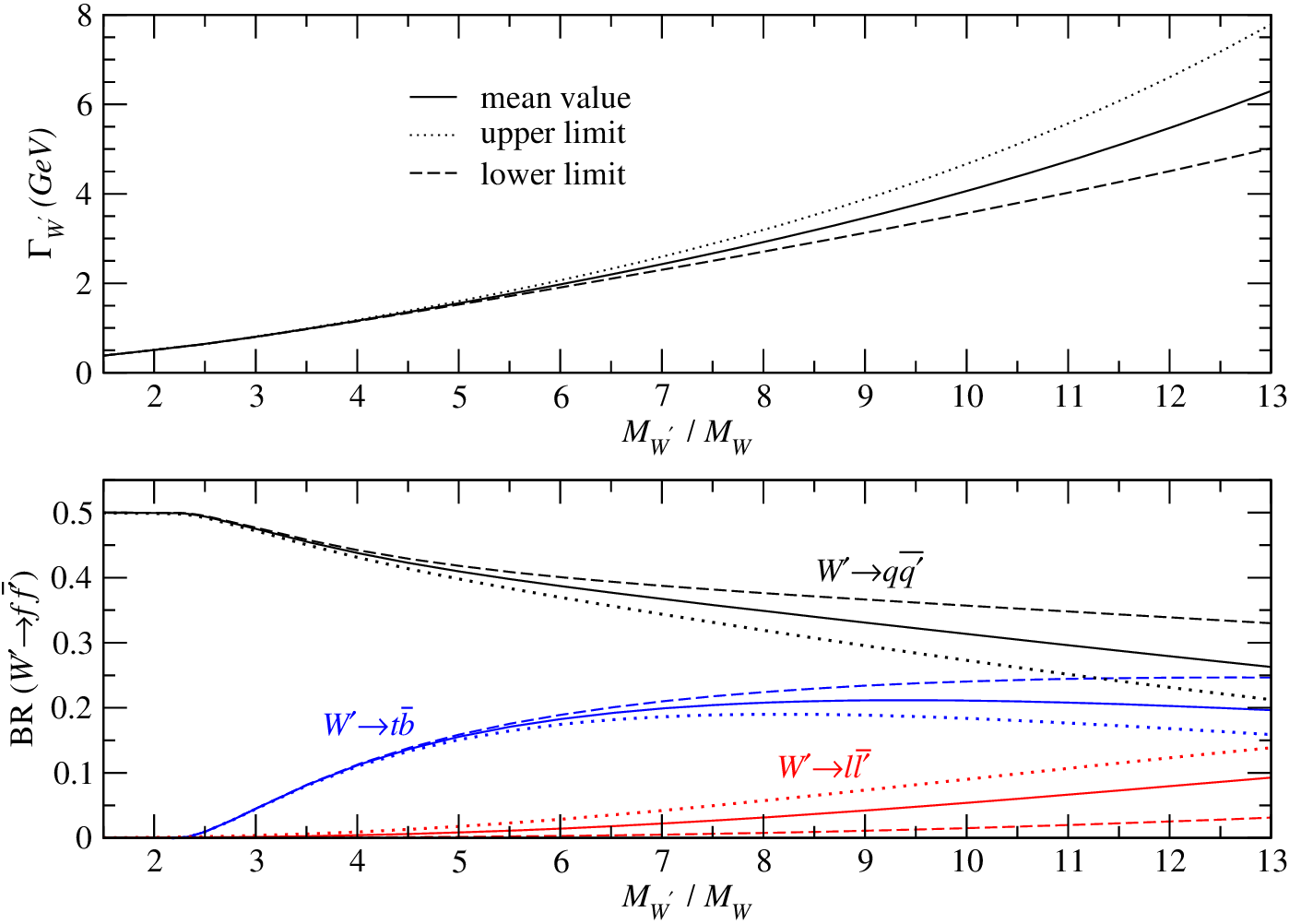}
\caption{(a) The total decay width and (b) the decay branching ratios
of the $W'$ boson as functions of its mass $M_{W'}$.\label{fig:totwidth_br}}
\end{figure}
In Fig.~\ref{fig:totwidth_br} we present the total decay width of the 
$W'$ boson and its decay branching ratios (BR) as functions of $M_{W'}$.
Here, we have separated the light quark decay modes (dashed-line)
from the heavy quark ($tb$) mode (dotted-line). It clearly shows
that in the region of small $M_{W'}$ ($1.5\, M_{W}<M_{W'}<2.5\, M_{W}$)
the light quark decay mode dominates over the other modes. This is
due to the suppression of the $g_{W'\ell\ell}$, cf. Fig.~\ref{fig:coup-mwp}.
As a result, the detection of $W'$ through its leptonic decay in the 
small $M_{W'}$ region is more difficult to achieve and the current
experimental data cannot rule out the existence of this $W'$. In
the medium mass region, the heavy quark decay channel opens. As a
result, the decay branching ratio of the light quark mode decreases but
is still larger than the heavy quark mode. Both hadronic decay
modes become comparable with increasing $M_{W'}.$ Again, the leptonic
decay mode is negligible due to the suppression of 
$g_{W'\ell\ell}$. In the region of very heavy $W'$, say
$M_{W'}>10\, M_{W}$, the leptonic decay branching ratio becomes larger
because $g_{W'qq}$ and $g_{W'\ell\ell}$ are of the same order.

\subsection{Discovery potential in hadron collision}

In this study, we will examine the discovery potential of the $W'$ boson
of this model at the Fermilab Tevatron and CERN Large Hadron
Collider (LHC). Many direct searches for a $W'$ boson in its various
decay modes have been performed at the Tevatron and produced lower limits
on its mass. The leptonic decay mode is the best choice for
disentangling the $W'$ event from the copious QCD background. Searches
using the decay mode $W'\to e\nu$ exclude a $W'$ boson with mass
$<754\,{\rm GeV}$ at $95\%$ C.L.~\cite{wprime-en1,wprime-en2},
while similar searches considering the decay mode $W'\to\mu\nu$ have
excluded a $W'$ boson with mass $<660\,{\rm GeV}$ at 
$95\%$ C.L.~\cite{wprime-nu}.
Combining both leptonic channels, the most stringent limit was obtained, 
excluding
a $W'$ boson with mass $<768\,{\rm GeV}$ at $95\%$ C.L.~\cite{wprime-en2}.
These mass limits all assume that the new vector boson's couplings
to leptonic final states are as given by the Standard Model, which
predicts that the total width of the boson increases linearly with
its mass.  In addition to the leptonic mode, a search using the light quark
decay mode $W'\to q\bar{q^{\prime}}$ excludes a $W'$ boson in the range 
$300<M_{W'}<420\,{\rm GeV}$
at $95\%$ C.L.~\cite{wprime-qq}, while a search using the decay
mode $W'\to t\bar{b}$ excludes a $W'$ boson in the range $225<M_{W'}<536\,{\rm GeV}$
for $M_{W'}\gg m_{\nu_{R}}$ and $225<M_{W'}<566\,{\rm GeV}$ for
$M_{W'}<m_{\nu_{R}}$~\cite{wprime-tb}. 

At a hadron collider the $W'$ bosons are predominantly produced through the 
charge-current Drell-Yan process: \[
q\bar{q^{\prime}}\to W'^{+}\to f\bar{f^{\prime}},\]
where $q$ and $q^{\prime}$ denote the light up-type quarks ($q=u,\, c$)
and down-type quarks ($q^{\prime}=d,\, s$) respectively. The total
cross section for this process at a hadron collider is
\begin{equation}
\sigma\left(P_{1}P_{2}\to f\bar{f^{\prime}}\right)=\sum_{q,\bar{q^{\prime}}}\int{\rm d}x_{1}{\rm d}x_{2}\left[f_{q/P_{1}}\left(x_{1},\mu\right)f_{\bar{q^{\prime}}/P_{2}}\left(x_{2},\mu\right)\hat{\sigma}\left(q\bar{q^{\prime}}\to f\bar{f^{\prime}}\right)+\left(x_{1}\leftrightarrow x_{2}\right)\right],
\end{equation}
where $P_{1}$, $P_{2}$ represent the hadronic initial state, $f_{q/P}(x,\mu)$
is the parton distribution function (PDF). We take the factorization
scale ($\mu$) to be the invariant mass of the constituent process in our numerical
calculation. The parton-level cross section $\hat{\sigma}$ is given by
\begin{equation}
\hat{\sigma}=\frac{1}{2\hat{s}}\int{\rm d}\Pi_{2}\sum_{\substack{{\rm spin}\\{\rm color}}}\overline{\left|\mathcal{M}\left(q\bar{q^{\prime}}\to f\bar{f^{\prime}}\right)\right|^{2}},\label{eq:partonlevel_xsec}
\end{equation}
where the bar over the $\left|\mathcal{M}\right|^{2}$ denotes averaging
over the initial-state spin and color, ${\rm d}\Pi_{2}$ represents
2-body final-state phase space, and the squared matrix element reads
\begin{equation}
\overline{\left|\mathcal{M}\right|^{2}}=\frac{N_{C}^{\, f}}{12}\frac{g_{L}^{4}\left|V_{ud}\right|^{2}}{64\pi^{2}\hat{s}}\frac{\hat{u}^{2}}{\left(\hat{s}-M_{W'}^{2}\right)^{2}+\left(M_{W'}\Gamma_{W'}\right)^{2}}g_{W'qq}^{2}g_{W'\ell\ell}^{2},
\end{equation}
where the explicit factor $1/12$ results from the average over the
quark spins and colors, and $N_{C}^{\, f}$ is the number of color
state of decay products:
\begin{equation}
N_{C}^{\, f}=\left\{ \begin{array}{cc}
1 & f=\ell,\\
3 & f=q.\end{array}\right.
\end{equation}
Here, the Mandelstam variables are defined by
\begin{eqnarray}
\hat{s}=\left(p_{u}+p_{d}\right)^{2}, & \hat{t}=\left(p_{d}-p_{\ell}\right)^{2}, & \hat{u}=\left(p_{u}-p_{\ell}\right)^{2},
\end{eqnarray}
where $p_{i}$ denotes the momentum of particle $i$. 

\begin{figure}[t]
\centering
\includegraphics[clip,scale=0.65]{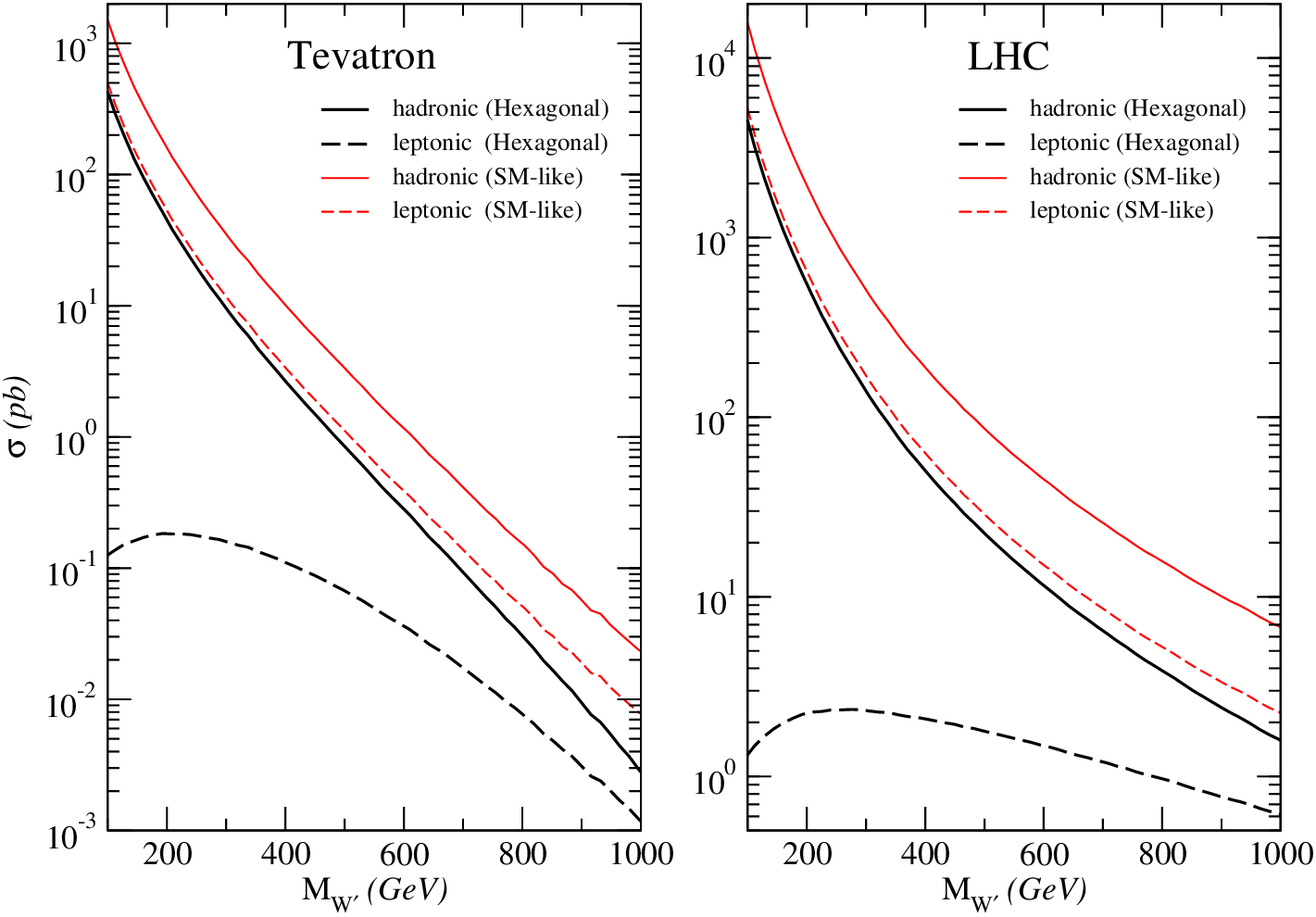}
\caption{Inclusive cross sections of $W'$ production and decay through 
$q\bar{q^{\prime}}\to W'\to f \bar f'$
as functions of $M_{W'}$ at the Tevatron and the LHC. \label{fig:inclusivexsec}}
\end{figure}

In Fig.~\ref{fig:inclusivexsec} we present the inclusive cross sections
of $W'$ production and decay through the process $u\bar{d}\to W'\to f\bar{f^{\prime}}$ at the Tevatron and the 
LHC. For comparison, we also present the inclusive cross sections of
the same process with the assumption that all the couplings are as in the 
Standard Model. For our numerical calculation, we use the leading-order parton
distribution function set CTEQ6L~\cite{cteq6}. The value of the
relevant electroweak parameters are $\alpha=1/137.0459895$, $G_{\mu}=1.16637\times10^{-5}\,{\rm GeV^{-2}}$,
$m_{t}=178\,{\rm GeV}$, $M_{W}=80.33\,{\rm GeV}$, $M_{Z}=91.1867\,{\rm GeV}$,
and $\sin^{2}\theta_{W}=0.231$. Thus, the square of the weak gauge
coupling is $g^{2}=4\sqrt{2}M_{W}^{2}G_{\mu}$. Here, we focus our
attention on the positively charged $W'$ boson only. 

Due to the suppression of the effective couplings ($g_{W'qq}$ and $g_{W'\ell\ell}$)
compared to those of the Standard Model, the inclusive cross section 
predicted by this model is much smaller than that of the SM, thereby shifting
the limits of  $M_{W'}$  to lower values. For example,
$W'$ couples to quarks with a suppression factor of $g_{W'qq}=g_{A}/g_{B}\simeq0.43$,
hence its production cross section will be a factor of 5 smaller
than expected for a corresponding gauge boson of the same mass in the SM.  
After using the kinematics cuts listed in Ref.~{\cite{wprime-qq}} for the 
light-quark mode, 
we compare the $W'$ cross section of our model with present data and 
conclude that $M_{W'}$ should be larger than $310~\rm{GeV}$. 
The leptonic mode is suppressed so much in our model that the $W'$ boson 
satifies all
the current experimental constraints, but it also means that one cannot detect 
this extra vector boson in this mode in the future.
At the LHC, as shown in Fig.~{\ref{fig:inclusivexsec}}, the production of $W'$ boson 
with its subsequent decay can be observed by studying events with two hard jets.  
Again, the leptonic decay mode is not very competitive. 
Detailed ananlysis of these two modes together with various backgrounds will be presented elsewhere.
We note also that the hadronic decay channel exhibits a completely
different behavior from the leptonic decay channel, especially for
a light $W'$ boson. This is a consequence of the difference
between the effective coupling strengths (cf. Fig.~\ref{fig:coup-mwp}),
and can be explained as follows.  Since the width of the $W'$ boson is 
very small compared to its mass, we can write the parton-level cross section
$\hat{\sigma}$ in Eq.~(\ref{eq:partonlevel_xsec}) as
\begin{equation}
\hat{\sigma}\left(q\bar{q^{\prime}}\to f\bar{f^{\prime}}\right)
=\hat{\sigma}\left(q\bar{q^{\prime}}\to W'^{+}\right)\times 
Br\left(W'^{+}\to f\bar{f^{\prime}}\right)
\end{equation}
under the narrow-width approximation. As an $s$-channel process, the
cross section $\hat{\sigma}(q\bar{q^{\prime}}\to W'^{+})$ drops off
rapidly with increasing $\hat{s}$ as 
$\hat{\sigma}\propto 1/{\hat{s}}$.
On the other hand, due to the large suppression of $g_{W'\ell\ell}$, 
the decay branching ratio of $W'\to\ell\bar{\ell^{\prime}}$ is
very tiny when $M_{W'}/M_{W}\leq5$ and increases with increasing
$M_{W'}$. These two effects compete with each other and leave the
bump in the inclusive cross section (cf. bold dashed curve in Fig.~\ref{fig:inclusivexsec}).

Since $(W'^{+},Z',W'^{-})$ is a triplet under $SU(2)_{L}$, $Z'$
has the same mass as $W'$ and the same couplings to quarks and leptons,
assuming no mixing with the SM gauge bosons.  As usual, one can use the leptonic decay 
mode to distinguish $W'$ from $Z'$. The $W'$ boson decays into one charged lepton 
and  one neutrino which has the collider signature of a charged lepton plus 
missing energy,
while the $Z'$ boson decays into two detectable charged leptons. In our model, however, we have to use 
the hadronic decay mode to detect these extra vector bosons, due to the suppresssion of the leptonic
decay mode discussed above. 
As far as the light-quark mode is concerned, both $W'$ and $Z'$ will have the collider signature
of two hard jets. Since both $W'$ and $Z'$ couple to quarks via  the left-handed gauge interaction, 
the two hard jets in the final state will have exactly the same kinematics distributions, 
it is thus impossible to distinguish one from the other.  On the other hand, one can easily separate them by
using the heavy-quark mode. For example, the $W'$ boson will decay into a $t \bar b$ pair with the top
quark subsequently decaying  into $\ell b\nu$ while the $Z'$ will decay into a $t\bar{t}$ pair with
the top-quark pair subsequently decaying into $\ell \bar \ell b \bar{b} \nu \bar{\nu}$. 

\section{Conclusion}
In this paper we have proposed a supersymmetric gauge extension of the 
Standard Model, where $SU(2)_L$ is enlarged to $SU(2)_{AL} \times SU(2)_{BL}$ 
at the TeV scale.  This model is motivated by (1) the possibility of 
$SU(3)^6$ hexagonal unification and (2) the possibility of small deviations 
from quark-lepton universality as allowed by neutron decay.

The distinguishing feature of our model is that quarks couple to 
$SU(2)_{AL}$ while leptons couple to a linear combination of $SU(2)_{AL}$ and 
$SU(2)_{BL}$ with mixing angle $\theta$.  The gauge couplings $g_A$ and 
$g_B$ are fixed from $SU(3)^6$ unification, and the mass of the 
$(W'^+,Z',W'^-)$ $SU(2)_L$ triplet is related to the angle $\theta$ from 
neutron decay.  We have discussed in this paper the possible production 
and decay of this new $W'$ boson. 
Using present Tevatron data, we set the lower limit of 310 GeV on $M_{W'}$ 
through its possible decay into quarks. [The leptonic mode turns out to be 
very much suppressed.]  Since $M_{W'}$ is expected to be no more than a few 
times $M_W$ in this particular theoretical context, it should become 
observable at the LHC.

\newpage
\section*{Acknowledgements}

This work was supported in part by the U.~S.~Department of Energy
under Grant No. DE-FG03-94ER40837.  GR thanks the Physics Department of 
the University of California, Riverside for hospitality, and acknowledges 
his support as a Raja Ramanna Fellow of the Department of Atomic Energy, 
Government of India.

\end{document}